\documentclass[reprint,aps,pre,twocolumn,showpacs]{revtex4-1}
\usepackage{mathtools}
\usepackage{amsmath, amsthm, amssymb}
\usepackage{graphicx}

\usepackage[normalem]{ulem}
\usepackage{xcolor}
\usepackage{hyperref}
\hypersetup{
  colorlinks = true
}

\def\pa{\partial}
\def\instD{{\mathcal P}_{\text{ex}}(\tSpec)}

\def\FI{\mathcal I}

\def\be{\begin{equation}}
\def\ee{\end{equation}}
\def\ba{\begin{align}}
\def\ea{\end{align}}
\def\s{\section}
\def\ss{\subsection}

\def\md{\mathrm{d}}
\def\f{\frac}
\def\l{\left}
\def\r{\right}

\def\Peq{P_{\textrm{eq}}}

\def\kB{k_{\text{B}}}
\def\kT{\kB T}
\def\la{\langle}
\def\ra{\rangle}
\def\kC{k_{\textrm{ch}}}
\def\kCL{\kC^{\textrm L}}
\def\kCR{\kC^{\textrm R}}
\def\kCLR{\kC^{\textrm{L/R}}}
\def\sech{\text{sech}}
\def\Fo{\rm F_{\textrm o}}
\def\F1{\rm F_1}
\def\kWL{\kW^{\textrm L}}
\def\kWR{\kW^{\textrm R}}
\def\kWLR{\kW^{\textrm{L/R}}}

\def\bDEdM{\beta \Delta E^{\ddagger}_{\textrm{m}}}
\def\eOffset{\Delta E}
\def\tI{t_{\textrm i}}
\def\tF{t_{\textrm f}}
\def\Em{E_{\textrm{m}}}
\def\Es{E_{\textrm{s}}}
\def\Wex{W_{\textrm{ex}}}

\newcommand{\sfrac}[2]{\ensuremath{\mathchoice{\tfrac{#1}{#2}}{\tfrac{#1}{#2}}{\frac{#1}{#2}}{\frac{#1}{#2}}}}
\newcommand{\half}{\sfrac{1}{2}}

\def\kS{k_{\textrm{s}}}
\def\kW{k_{\textrm m}}
\def\xS{x_{\textrm{s}}}
\def\xMin{x_{\textrm m}}
\def\tSpec{t'}

\begin{document}

\title{Thermodynamic geometry of minimum-dissipation driven barrier crossing}
\date{\today}

\author{David A.\ Sivak}
\email{dsivak@sfu.ca}
\affiliation{Dept. of Physics, Simon Fraser University, Burnaby, British Columbia V5A 1S6, Canada}
\author{Gavin E.\ Crooks}
\affiliation{Molecular Biophysics Division, Lawrence Berkeley National Laboratory, Berkeley, CA 94720, USA}
\affiliation{Kavli Energy NanoSciences Institute at Berkeley, CA 94720, USA}

\begin{abstract}
We explore the thermodynamic geometry of a simple system that models the bistable dynamics of nucleic acid hairpins in single molecule force-extension experiments. Near equilibrium, optimal (minimum-dissipation) driving protocols are governed by a generalized linear response friction coefficient. 
Our analysis demonstrates that the friction coefficient of the driving protocols is sharply peaked at the interface between metastable regions, which leads to minimum-dissipation protocols that drive rapidly within a metastable basin, but then linger longest at the interface, giving thermal fluctuations maximal time to kick the system over the barrier. Intuitively, the same principle applies generically in free energy estimation (both in steered molecular dynamics simulations and in single-molecule experiments), provides a design principle for the construction of thermodynamically efficient coupling between stochastic objects, and makes a prediction regarding the construction of evolved biomolecular motors.
\end{abstract}

\maketitle

\s{Introduction}
Molecular machines built from protein complexes are critical players in numerous cellular processes which convert between different forms of energy, from muscle contraction to intracellular transport of organelles and chromosomes to crawling or swimming~\cite{Mavroidis:2004:AnnRevBiomedEng}.  
Central to molecular machine function is their thermodynamic efficiency, that is, their ability to translate free energy input into useful work without losing too much energy in the form of heat dissipated into the environment. Given high turnover and the costs associated with energy dissipation, it seems plausible that evolution has sculpted these machines to avoid needlessly wasting energy.  

Indeed, several biomolecular machines (perhaps most notably the $\F1$ subunit of ATP synthase) have been shown to have near-perfect efficiency at stall force or torque~\cite{Yasuda:1998:Cell}.  
However, machines that must turn over on a timescale of tens to hundreds of milliseconds do not operate near the slow, quasistatic limit~\cite{Etzold:1997:EurJBiochem}. It behooves us to ask: What are the limits of the energetic efficiency of these fluctuating soft-matter objects when they operate rapidly and hence are driven far from equilibrium?  
Furthermore, what mechanical manipulations of these machines or within these machines attain these limits?

Thus there is a growing interest in general methods for finding efficient protocols to drive nonequilibrium processes. 
Conceptually, such a method would provide a framework for understanding machine behavior, and for predicting the interactions between components in biological systems (e.g., the $\Fo$ and $\F1$ subunits within ATP synthase~\cite{Okuno:2011:Biochem}) that have been evolutionarily tuned to be energetically efficient~\cite{Yasuda:1998:Cell}.
Practically, single-molecule force-extension experiments~\cite{Gore:2003:PNAS} and steered molecular dynamics simulations for measuring free energy differences require less repetitions for a given confidence interval when they dissipate less energy~\cite{Crooks:2007:PhysRevLett,Maragakis:2008:JChemPhys,Shenfeld:2009:PhysRevE, Grosse:2013:NIPS}, 
so methods that identify low-dissipation protocols promise to improve the efficiency of both experiments and numerical simulations.  
Such progress can also help guide the design of synthetic molecular machines~\cite{Hess:2011:AnnRevBiomedEng}, for example, to improve artificial photosynthesis~\cite{Moore:2011:AnnRevCondMattPhys}. 

Recent theoretical advances in the field of nanoscale nonequilibrium thermodynamics have provided tools to understand the nonequilibrium processes that these molecular motors perform. 
Exact results exist for some simple models~\cite{Schmiedl:2007:PhysRevLett,GomezMarin:2008:JChemPhys}; nevertheless, all but the simplest models, and indeed any multi-dimensional protocol, remain beyond the scope of exact analysis.
We have recently developed a linear response framework that, through a generalized friction coefficient in control parameter space, 
gives a near-equilibrium approximation for the system response to nonequilibrium driving and hence an estimate for the average excess work exerted in rapid driving of an arbitrary number of control parameters~\cite{Sivak:2012:PhysRevLett:b,Zulkowski:2012:PhysRevE,Zulkowski:2013:PLoSONE,Zulkowski:2014:PhysRevE,Zulkowski:2015:PhysRevE,Rotskoff:2015:PhysRevE}.  
This friction coefficient reports on the resistance the system puts up to rapid changes in the control parameter. 

In this work we explore the implications of this theoretical framework for a model system of wide applicability throughout biophysics and soft matter: a continuous analog of a two-state system, a one-dimensional system with two metastable mesostates separated by an energetic barrier, driven by an additional time-dependent quadratic potential.  
This most obviously forms a model for the force-induced unfolding using optical tweezers or atomic force microscopy (AFM) of a DNA or RNA hairpin~\cite{Strick:2003:RepProgPhys}.  
To further the goals of optimizing and designing efficient finite-time microscopic nonequilibrium processes, we examine this generalized friction coefficient and the resulting optimal protocols.

\s{Theoretical framework}
This section largely summarizes the original linear response derivation in Ref.~\cite{Sivak:2012:PhysRevLett:b} of Eq.~\eqref{eq:punchline} in the current paper; see Refs.~\cite{Mandal:2016:JStatMech,RotskoffCrooksEVE} for alternative routes to the same equation.
A physical system at thermal equilibrium with a heat reservoir at temperature $T$ is distributed over microstates $x$ according to the canonical ensemble
\be
\label{canonical} 
\pi( x \mid \lambda) \equiv \exp{ \beta \bigl[ F(\boldsymbol \lambda) - E(x,\boldsymbol \lambda) \bigr] } \ ,
\ee
where $\beta = (\kT)^{-1}$ is the inverse temperature in natural units, $E(x,\boldsymbol \lambda)$ is the system energy as a function of the microstate $x$ and a collection of experimentally controllable parameters $\boldsymbol \lambda$, and $F(\boldsymbol \lambda) \equiv -\kT \ln \int \md x \, \exp[-\beta E(x,\boldsymbol \lambda)]$ is the free energy.

The instantaneous rate of energy flow into the system
\be
\label{natenergy} 
\frac{\md}{\md t} E(x, \boldsymbol \lambda) = \frac{\md x^{T}}{\md t} \cdot  \frac{\partial}{\partial x}E(x, \boldsymbol \lambda) + \frac{\md \boldsymbol \lambda^{T}}{\md t} \cdot  \frac{\partial}{\partial \boldsymbol \lambda}E(x, \boldsymbol \lambda) \ ,
\ee
naturally splits into energy flow due to system fluctuations (heat flow, the first term on RHS), 
and energy flow due to changes of the external parameters (work, second term on RHS)~\cite{Jarzynski:2011:AnnRevCondMattPhys}.

The excess power exerted at time $\tSpec$ by the external agent on the system (averaged over the ensemble of system responses), over and above the average power on an equilibrated system, is
\be
\label{expower} 
\mathcal{P}_{\textrm{ex}}(\tSpec) \equiv - \l[ \frac{ \md \boldsymbol \lambda^{T} }{\md t} \r]_{\tSpec} \cdot \l\la \triangle \boldsymbol f \r\ra_{ \boldsymbol \Lambda} \ .
\ee
Here angled brackets with subscript $\boldsymbol \Lambda$ indicate a nonequilibrium average dependent on the protocol $\boldsymbol\Lambda$, the time course of the control parameter $\lambda$. 
$\boldsymbol f \equiv -\frac{ \partial \l( \beta E \r)}{\partial \boldsymbol \lambda}$ are the forces conjugate to the control parameters $\boldsymbol \lambda$, and $\triangle \boldsymbol f (\tSpec) \equiv \boldsymbol f(\tSpec) - \l\la \boldsymbol f \r\ra_{\boldsymbol \lambda(\tSpec) }$ is the deviation of ${\bf f}(\tSpec)$ from $\l\la \boldsymbol f \r\ra_{\boldsymbol \lambda(\tSpec)}$, its equilibrium value at the current control parameter.
The Second Law of thermodynamics imposes non-negativity on this average excess work for any protocol.  

Protocols that change in response to measurements of the system can seemingly evade such limits, although the subtle but inescapable thermodynamic costs of information processing means there is no free lunch~\cite{Parrondo:2015:NatPhys}. Such generalizations are beyond the scope of this paper, where we restrict our attention to protocols that are specified beforehand, with no feedback based on the intermediate state of the system.

For twice-differentiable protocols, applying linear response theory~\cite{Zwanzig:2001} gives an average excess power~\cite{Sivak:2012:PhysRevLett:b}
\be
\mathcal{P}_{\textrm{ex}}(\tSpec) \approx \l[ \frac{ \md \boldsymbol \lambda^{T} }{\md t} \r]_{\tSpec} \cdot \zeta\bigl(\boldsymbol \lambda(\tSpec) \bigr) \cdot \l[ \frac{ \md \boldsymbol \lambda }{\md t}  \r]_{\tSpec} \ ,
\label{eq:punchline}
\ee
for the generalized friction tensor
\be
\label{newfrictensor} 
\zeta_{ij}\bigl(\boldsymbol \lambda(\tSpec)\bigr) \equiv  \beta \int_{0}^{\infty} \md t'' \bigl\langle \delta f_{j}(0) \ \delta f_{i}(t'') \bigr\rangle_{\boldsymbol \lambda(\tSpec)}\ .
\ee
Here $\bigl\langle \delta f_{j}(0)\ \delta f_{i}(t'') \bigr\rangle_{\boldsymbol \lambda(\tSpec)}$ is the force autocorrelation function defined in terms of equilibrium fluctuations $\delta f_i(t) \equiv f_i(t) - \la f_i \ra_{\boldsymbol \lambda(\tSpec)}$. 
When all conjugate forces are even under reversal of momenta, this friction tensor $\zeta$ is symmetric, positive semidefinite, and smoothly varying except at macroscopic phase transitions, and thus induces a Riemannian geometry on the space of thermodynamic states~\cite{McCleary:1995}.
Intuitively, a system that relaxes quickly to equilibrium, compared to the rapidity of perturbation, is sufficient (though not necessary~\cite{Mazonka:1999:arxiv}) for the linear response approximation to hold.

For a single control parameter $\lambda$, this simplifies to
\begin{subequations}
\begin{align}
\instD &= \zeta\bigl(\lambda(\tSpec)\bigr) \l( \f{\md\lambda}{\md t} \r)^2 
\\
\zeta\bigl(\lambda(\tSpec)\bigr) &= \beta \ \tau\bigl(\lambda(\tSpec)\bigr)\ \la \delta f^2 \ra_{\lambda(\tSpec)} \ ,
\label{eq:TA}
\end{align}
\end{subequations}
for the force variance $\la \delta f^2\ra_{\lambda(\tSpec)} = \la \delta f(0) \ \delta f(0) \ra_{\lambda(\tSpec)}$ and the integral force relaxation time~\cite{Garanin:1996:PhysRevE}
\be
\tau\bigl(\lambda(\tSpec)\bigr) \equiv \int_0^{\infty} \md t''\ \f{\bigl\langle \delta f(0)\ \delta f(t'') \bigr\ra_{\lambda(\tSpec)}}{\bigl\la \delta f^2 \bigr\ra_{\lambda(\tSpec)}} \ .
\ee

\s{Model system}
\begin{figure}[t]
\centering
\includegraphics[width=\columnwidth]{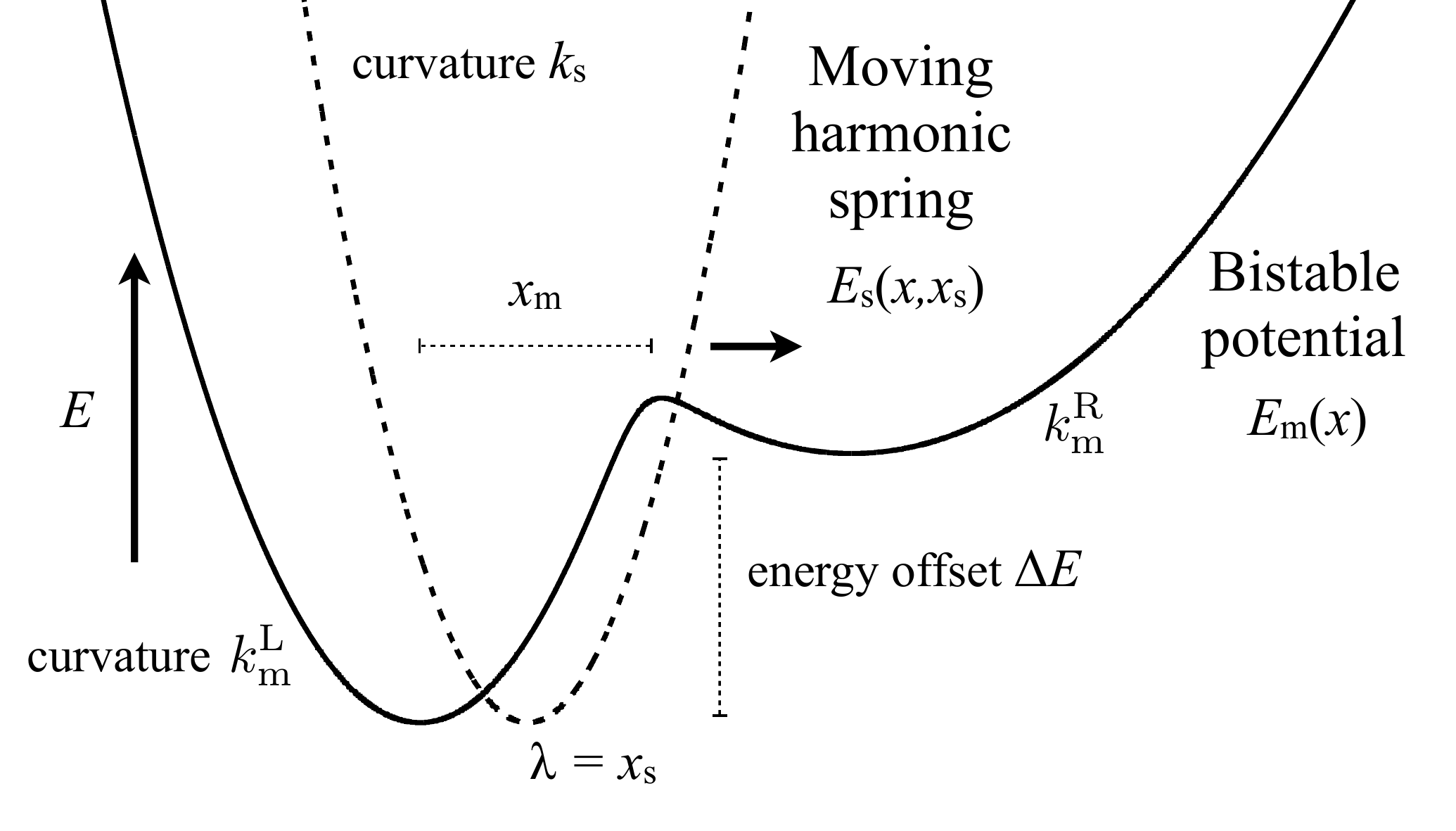}
\caption{The total potential $E(x,\xS)$ is the sum of a time-independent bistable potential $\Em(x)$ (solid curve) and a harmonic potential $\Es(x,\xS)$ (dashed curve) whose location depends on time through the control parameter $\lambda=\xS$, the trap minimum.}
\label{fig:Pot}
\end{figure}

We simulate a single particle diffusing over a one-dimensional energy profile $E(x, \xS) = \Es(x,\xS) + \Em(x)$ composed of two components (Fig.~\ref{fig:Pot}). The bistable molecular potential
\begin{align}
\Em&(x) = \\
&-\kT\ln\l\{e^{-\half \beta \kWL (x + \xMin)^2} 
 + e^{-\beta\l[\half \kWR (x - \xMin)^2 + \eOffset\r]} \r\}
\notag
\end{align}
is motivated by the statistical mechanics of a system with two metastable ensembles of conformational states, such as a two-state biomacromolecule with folded- and unfolded-state ensembles (e.g., an DNA or RNA hairpin). 
The two conformational states each induce a quadratic potential, centered on $x=\xMin$ and $-\xMin$ respectively. 
The right (unfolded) state has an energy offset $\eOffset$ from the left (folded) state. The specific form of the potential gives the free energy of the particle assuming that at each position coordinate it fluctuates between the two conformational states, with residence probabilities given by the Boltzmann weights of each conformational state at that particular position. 

\begin{figure}[t]
\centering
\includegraphics[width=\columnwidth]{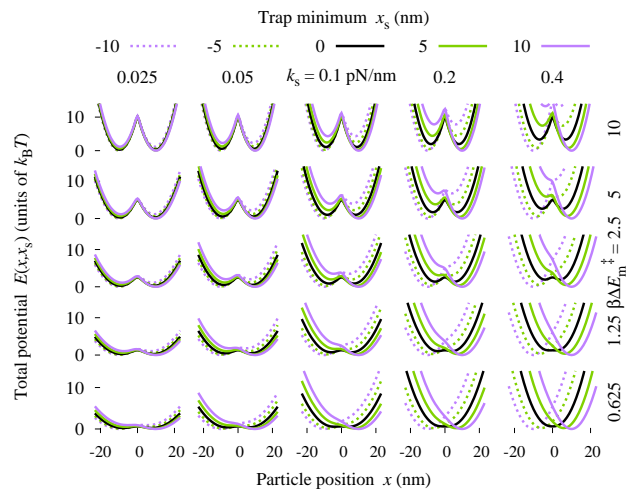}
\caption{Total potential energy landscape $E(x,\xS)$ [bistable potential $\Em(x)$ plus quadratic potential $\Es(x,\xS)$] as a function of particle position $x$. Different curves within a sub-plot have different trap minima $\xS$. Ascending rows have higher molecular barrier heights $\bDEdM$. Columns to the right have higher trap strengths $\kS$.}
\label{fig:potentials}
\end{figure}

The harmonic spring potential $\Es(x,\xS) = \half \kS(x-\xS)^2$, with time-dependent minimum $\xS$ and spring constant $\kS$, represents mechanical manipulation by optical tweezers or AFM, both hereafter generically referred to as a `trap.' 
The single control parameter $\lambda = \xS$  (the location of the minimum of the harmonic potential) represents the preferred separation imposed by the trap, for example the distance between foci of two optical traps, or between an immobilized surface and AFM cantilever.  
During a typical force-extension experiment this minimal-energy separation is increased to unfold the macromolecule (in our model pulling the particle from the left to the right basin) or decreased to refold the macromolecule. 
For varying trap minimum the particle experiences a varying total potential (Fig.~\ref{fig:potentials}).

Straightforward calculus leads to the equilibrium probability distribution
\begin{align}
&\pi(x \mid \xS) = \\
&\f{1}{Z}e^{-\half \beta \kS(x-\xS)^2}\l( e^{-\half\beta \kWL(x+\xMin)^2} + e^{-\beta\l[\eOffset + \half \kWR(x-\xMin)^2\r]} \r)\ \nonumber
\end{align}
for the partition function
\begin{align}
Z 
&= \sqrt{\f{2\pi}{\beta}} \l(\f{e^{-\half \beta \kCL (\xS+\xMin)^2}}{\sqrt{\kS+\kWL}} + \f{e^{-\beta\l[\eOffset + \half \kCR (\xS-\xMin)^2 \r]}}{\sqrt{\kS+\kWR}} \r) \ .
\end{align}
Here the characteristic spring constant 
\be
\kCLR \equiv [\kS^{-1} + (\kWLR)^{-1}]^{-1} = \f{\kS\kWLR}{\kS + \kWLR} 
\ee
is half the harmonic mean of the two spring constants $\kS$ and $\kWLR$.
To ease analytic interpretation and reduce the dimensionality of parameter space, we henceforth restrict our attention to basins of equal curvature ($\kWL=\kWR=\kW$).

For no energy offset ($\eOffset=0$), there are tractable expressions for the activation energies between the barrier and the metastable basins. For a significant energy barrier ($\half \kW \xMin^2 \gg \kT$), the molecular activation energy is approximately $\Delta E^{\ddagger}_{\textrm{m}} \equiv \Em^{\textrm{barrier}}(x) - \Em^{\textrm{min}}(x) \approx \tfrac{1}{2}\kW \xMin^2 - \kT \ln 2$.
The second term reflects the entropic benefit of equal accessibility to each conformational state.
When the trap minimum is halfway between the two basins at the molecular energy barrier (the hopping regime for single-molecule experiments), for steep wells ($2\kW\xMin^2/[1+\kS/\kW] \gg 1$) the total activation energy (including quadratic trap) is
\be
\Delta E^{\ddagger} \approx \f{\half \kW\xMin^2}{1+\f{\kS}{\kW}} - \kT\ln 2 \ .
\ee

We calculate the actual excess work using a dynamic programming algorithm~\cite{Sivak:2012:PhysRevLett:a} to dynamically propagate the nonequilibrium position distribution. To calculate the excess power from \eqref{eq:TA}, the control parameter velocity $\md \xS/\md t$ is dictated by the protocol, and the force variance $\la \delta f^2 \ra_{\xS}$ is analytically solvable for this model.  Previously, calculating the force relaxation time for this type of model~\cite{Bonanca:2014:JChemPhys} required numerical simulations.
However, a recent advance has analytically simplified the full friction coefficient for one-dimensional overdamped diffusive dynamics to~\cite{Zulkowski:2015:PhysRevE,Berezhkovskii:2011:JChemPhys}
\be
\zeta_{ij}(\boldsymbol \lambda) = \f{1}{D} \int_{-\infty}^{\infty} \md x \l[ \f{\pa_{\lambda^i} \Pi_{\textrm{eq}}(x,{\boldsymbol \lambda}) \ \pa_{\lambda^j} \Pi_{\textrm{eq}}(x,{\boldsymbol \lambda})}{\Peq(x,{\boldsymbol \lambda} )} \r] \ ,
\ee
requiring only the diffusion coefficient $D$ and 
the cumulative distribution function $\Pi_{\textrm{eq}}(x,{\boldsymbol \lambda})$.  

We explore this model by characterizing a parameter regime roughly corresponding to contemporary optical tweezer experiments on single nucleic-acid hairpins: diffusion coefficient $D = 0.44$ $\mu$m$^2$/s (dominated by the diffusivity of the micron-sized optical bead), pulling velocity $v = 100$ nm/s, distance $\xMin = 10$ nm from basin to barrier, trap stiffness $\kS = 0.025-0.4$ pN/nm, and molecular barrier height $\bDEdM = 0.625-10$ $\kT$.

\s{Results}

\begin{figure}[!tbp]
\centering
\includegraphics[width=\columnwidth]{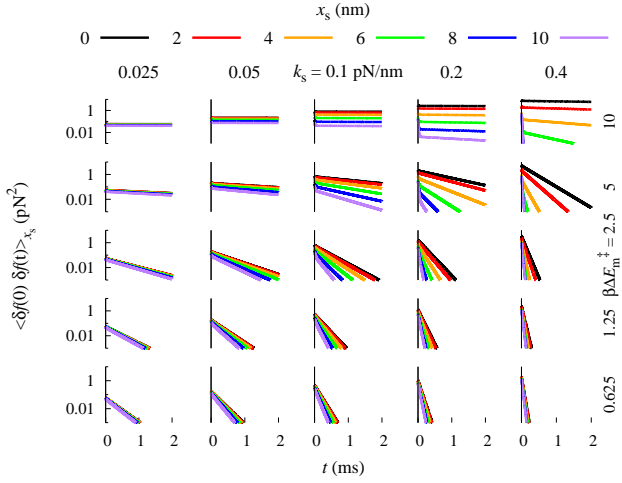}
\caption{Equilibrium force autocorrelation function $\langle \delta f(0) \delta f(t) \rangle_{\xS}$ for trap minimum $\xS$, with same variation of $\xS$, $\kS$, and $\bDEdM$ as in Fig.~\ref{fig:potentials}.}
\label{fig:autocorr}
\end{figure}

\ss{Friction coefficient}
The force autocorrelation function varies dramatically with varying trap stiffness $\kS$, molecular barrier height $\bDEdM$, and trap minimum $\xS$ (Fig.~\ref{fig:autocorr}).  

\begin{figure}[!tbp]
\centering
\includegraphics[width=\columnwidth]{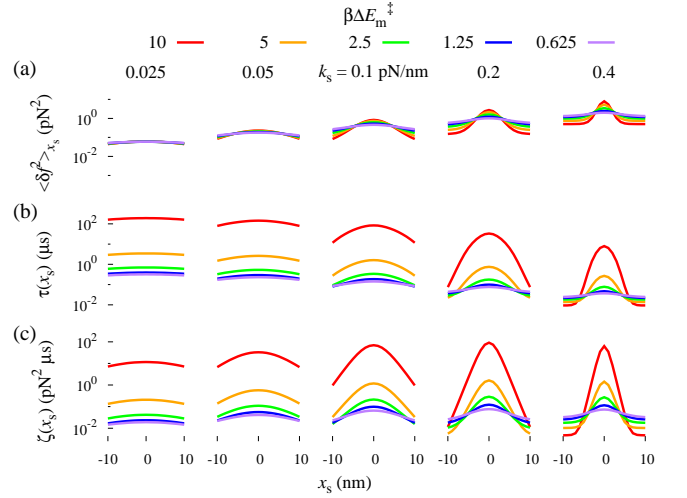}
\caption{(a) Force variance $\la \delta f^2 \ra_{\xS}$, (b) relaxation time $\tau(\xS)$, and (c) their product, the generalized friction coefficient $\zeta(\xS)$, across a range of trap minima $\xS$ (within a curve), for varying molecular barrier heights $\bDEdM$ (different curves within given sub-plot) and varying trap stiffness $\kS$ (left to right sub-plots).}
\label{fig:varTFric}
\end{figure}
The generalized friction coefficient [Fig.~\ref{fig:varTFric}(c)] can be decomposed [Eq.~\eqref{eq:TA}] into the force variance and the integral force relaxation time [Figs.~\ref{fig:varTFric}(a) and \ref{fig:varTFric}(b)], which are both higher when the time-dependent trap minimum is at the molecular energy barrier, giving an equilibrium distribution with significant probability on either side of the barrier.  Across the range of model parameters we explored, the force variance spans two orders of magnitude and the force relaxation time four orders of magnitude.  

The symmetric double-well potential has force variance
\begin{align}
&\la \delta f^2 \ra_{\xS} = \\
&\kT\l( \f{\kS}{1+\f{\kW}{\kS}} + \l\{\kC \xMin\ \sech[\beta(\kC\xMin \xS - \half\eOffset)]\r\}^2 \r) \ . \nonumber
\label{eq:forceVar}
\end{align}
The variance peaks in the transition region for trap minimum $\xS = \eOffset/(2\kC\xMin)$. 
Increasing $\beta \kC\xMin$ compresses the region of large variance. 
Changing the energy offset $\eOffset$ simply shifts the location of the maximal variance, and thus in the rest of this paper we set $\eOffset = 0$. 

The analytic expression for the relaxation time is sufficiently complicated to defy easy interpretation, but its numerical results show a similar qualitative pattern to the variance.
Thus the friction coefficient peaks at the transition region, where both the force variance and force relaxation time are maximized.
The peak friction coefficient value scales with both $\kS$ and $\bDEdM$.  

Far from the transition region (on either side) the total potential is essentially quadratic with effective curvature $\kS+\kW$, minimum energy at $x=(\kS\xS+\kW\xMin)/(\kS+\kW)$, and hence translation velocity $(\pa \xS/\pa t)/(1+\kW/\kS)$  of the energy minimum.  Analytic solutions are available for the position and work distributions~\cite{Mazonka:1999:arxiv}. 
The force variance and relaxation time are both constant,  
\begin{subequations}
\begin{align}
\la \delta f^2 \ra_{\xS\rightarrow\pm\infty} &= \f{\kT \kS}{1+\f{\kW}{\kS}} \\
\tau(\xS\rightarrow\pm\infty) &= \f{\kT}{D(\kS+\kW)} \ .
\end{align}
\end{subequations}
This produces a friction coefficient that is also constant far from the barrier,
\be
\zeta(\xS\rightarrow\pm\infty) = \f{\kT}{D(1+\f{\kW}{\kS})^2} \ .
\ee

The analysis inspiring this paper~\cite{Sivak:2012:PhysRevLett:b} was a microscopic and dynamical generalization of `thermodynamic length' ideas originally derived for macroscopic systems~\cite{Weinhold:1975:JChemPhys,Ruppeiner:1979:PhysRevA,Salamon:1983:PhysRevLett}.
In Appendix~\ref{sec:TDTL} we examine the central quantities of that framework in this tractable model system.

\begin{figure}[!tbp]
\centering
\includegraphics[width=\columnwidth]{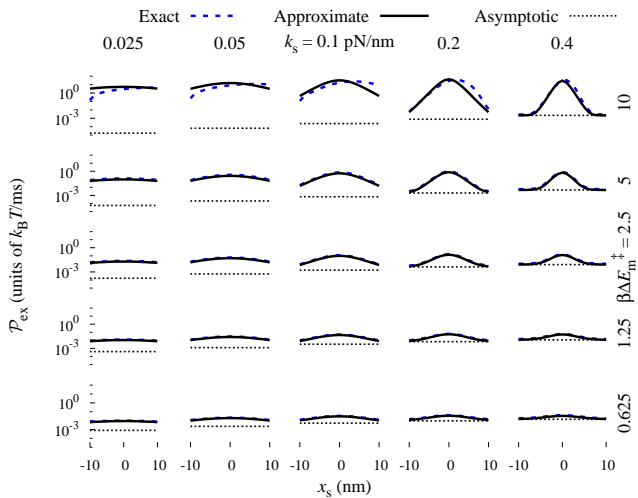}
\caption{Excess power ${\mathcal P}_{\textrm{ex}}$ as a function of control parameter $\xS$, calculated directly via numerical simulation (dashed blue curves) or estimated using the control parameter velocity and the cumulative distribution function form of the friction coefficient (solid black curves). Dotted black horizontal lines show asymptotic excess powers at $\xS = \pm \infty$.
Same variation of $\kS$ and $\bDEdM$ as in Fig.~\ref{fig:potentials}.
}
\label{fig:compDiss}
\end{figure}

\ss{Naive protocols}
Figure~\ref{fig:compDiss} shows the excess power for naive (constant-velocity) protocols, calculated directly from numerical Metropolis Monte Carlo simulations~\cite{Crooks:1999:PhysRevE} that do not assume linear response (dashed blue curves) and estimated analytically under the linear response approximation (solid black curves). When the molecular barrier height $\bDEdM$ is lower (shallower basins, lower sub-plots), the system remains closer to equilibrium and the approximation works well, reproducing very closely the exact excess power.  
Where the system is farther from equilibrium (higher barriers, upper sub-plots), the expression breaks down as the excess power becomes asymmetric. 
The approximation works quite well until reaching the middle transition region, when the trap minimum (and preponderance of the equilibrium probability distribution) crosses over to the right basin while the nonequilibrium probability density remains on the left side. 
Once the nonequilibrium density is pulled over to the right basin, the approximation once again captures the exact excess power. As the approximation involves only the current trap minimum~$\xS$ and its current velocity $\md \xS/\md t$, the approximation will never be accurate where, due to the history of the control parameter protocol, the nonequilibrium probability density is stuck in a qualitatively distinct region of state space.   

\ss{Optimal protocols}
Under the linear response approximation, the optimal (minimimum-dissipation) protocol proceeds such that the excess power is constant over the entire protocol, and thus the protocol velocity is proportional to the inverse square root of the friction coefficient~\cite{Salamon:1983:PhysRevLett, Crooks:2007:PhysRevLett, Sivak:2012:PhysRevLett:b}, 
\be
\f{\md \lambda^{\mathrlap{\text{opt}}} } {\md t}\quad \propto \zeta^{-1/2} \ .
\label{eq:optProt}
\ee
Under the linear response approximation, the shape of the optimal protocol is not a function of the allocated time interval. A shorter optimal protocol has a higher proportionality constant in Eq.~\eqref{eq:optProt} and hence produces a higher excess power, but the relative velocities at different points in the protocol remain unchanged.

\begin{figure}[!tbp]
\centering
\includegraphics[width=\columnwidth]{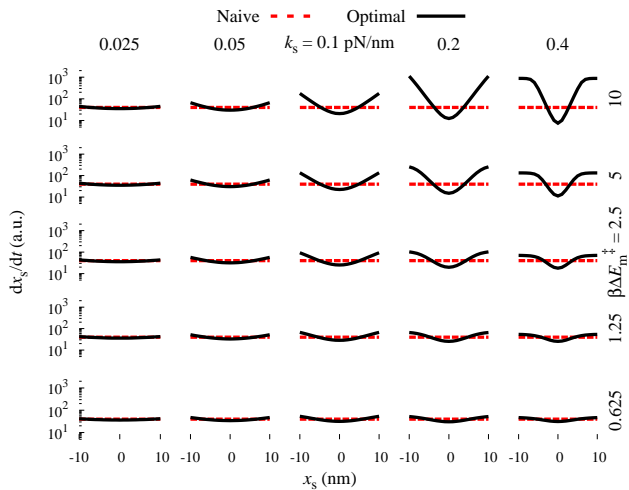}
\caption{Control parameter velocity $\md \xS/\md t$ (in arbitrary units) as a function of control parameter $\xS$, for naive constant-velocity protocols (red dashed lines) and optimal minimum-dissipation protocols under the linear response approximation (solid black curves). Same variation of $\kS$ and $\bDEdM$ as in Fig.~\ref{fig:potentials}.}
\label{fig:optVel}
\end{figure}

\begin{figure}[!tbp]
\centering
\includegraphics[width=\columnwidth]{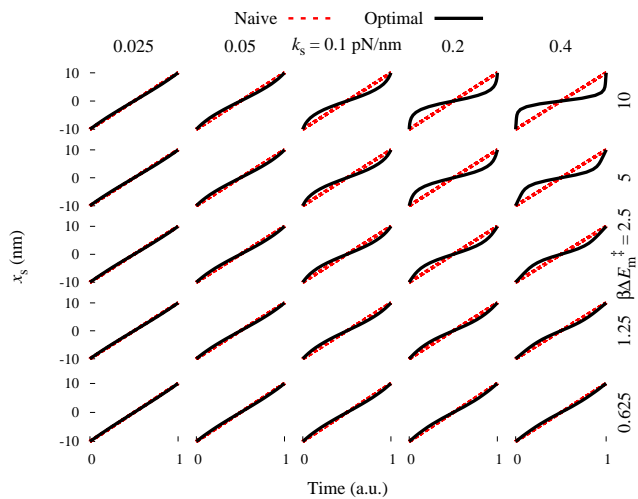}
\caption{Control parameter position $\xS$ as a function of time (in arbitrary units), for naive constant-velocity protocols (red dashed lines) and optimal minimum-dissipation protocols under the linear response approximation (solid black curves). Same variation of $\kS$ and $\bDEdM$ as in Fig.~\ref{fig:potentials}.}
\label{fig:OP}
\end{figure}

Given the variation in friction coefficient [Fig.~\ref{fig:varTFric}(a)], the optimal control parameter velocity (Fig.~\ref{fig:optVel}) can vary by orders of magnitude across a given protocol, leading to an optimal protocol that differs substantially from the naive constant-velocity protocol (Fig.~\ref{fig:OP}). 
Where the friction coefficient varies little (soft trap and small barrier, bottom left of Figs.~\ref{fig:optVel} and \ref{fig:OP}), the optimal minimum-dissipation protocol and naive constant-velocity protocol differ little. 

Where the friction coefficient varies by orders of magnitude across the protocol (stiff trap and large barrier, top right), the optimal protocol proceeds rapidly when the system relaxes quickly, far from the central transition region. Across the transition region the optimal protocol moves slowly to maximize the time spent in the hopping regime, giving thermal fluctuations as much time as possible to kick the system over the barrier without significant work input. Past the transition region, the optimal protocol again proceeds rapidly to the end.

\begin{figure}[!tbp]
\centering
\includegraphics[width=\columnwidth]{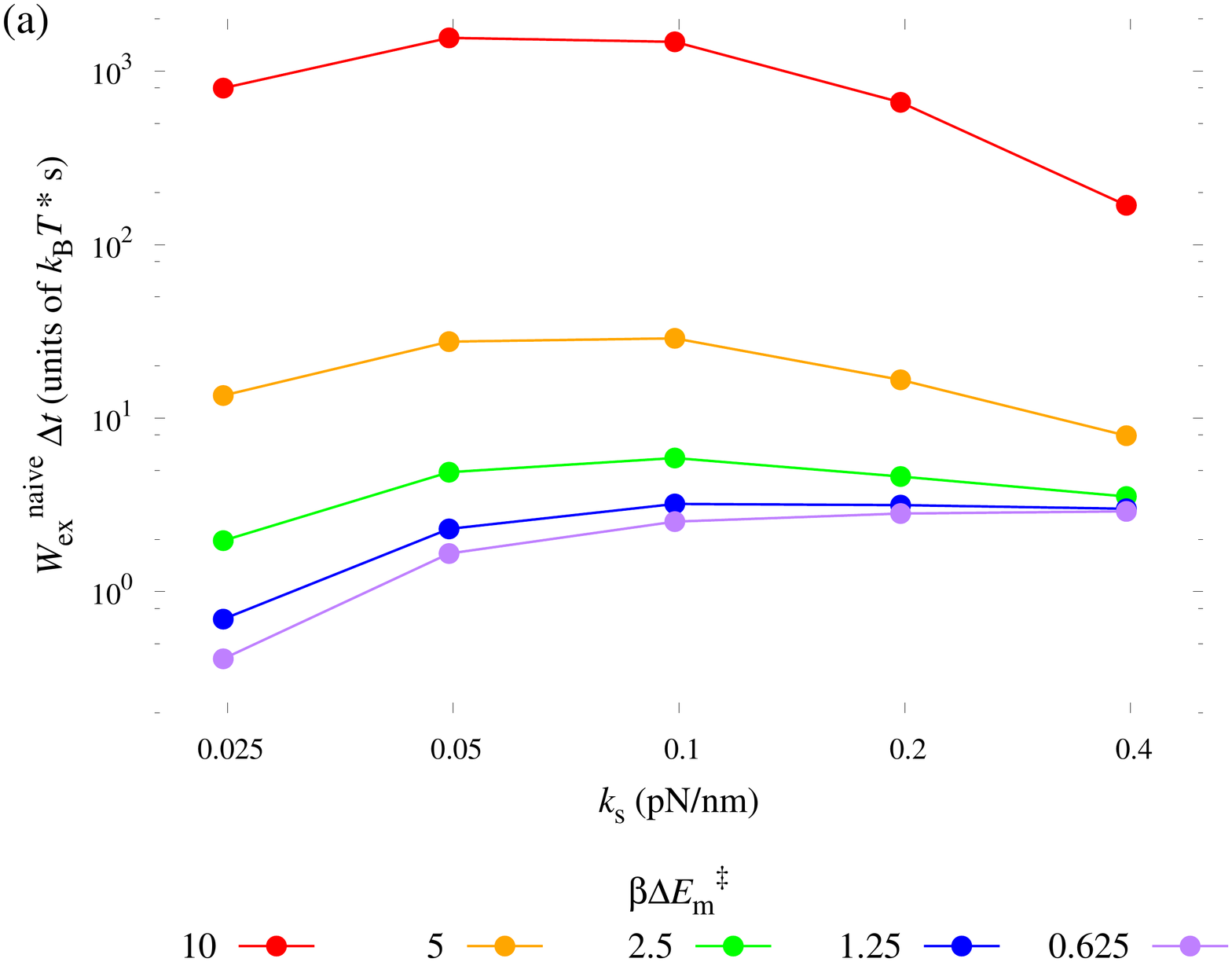}
\includegraphics[width=\columnwidth]{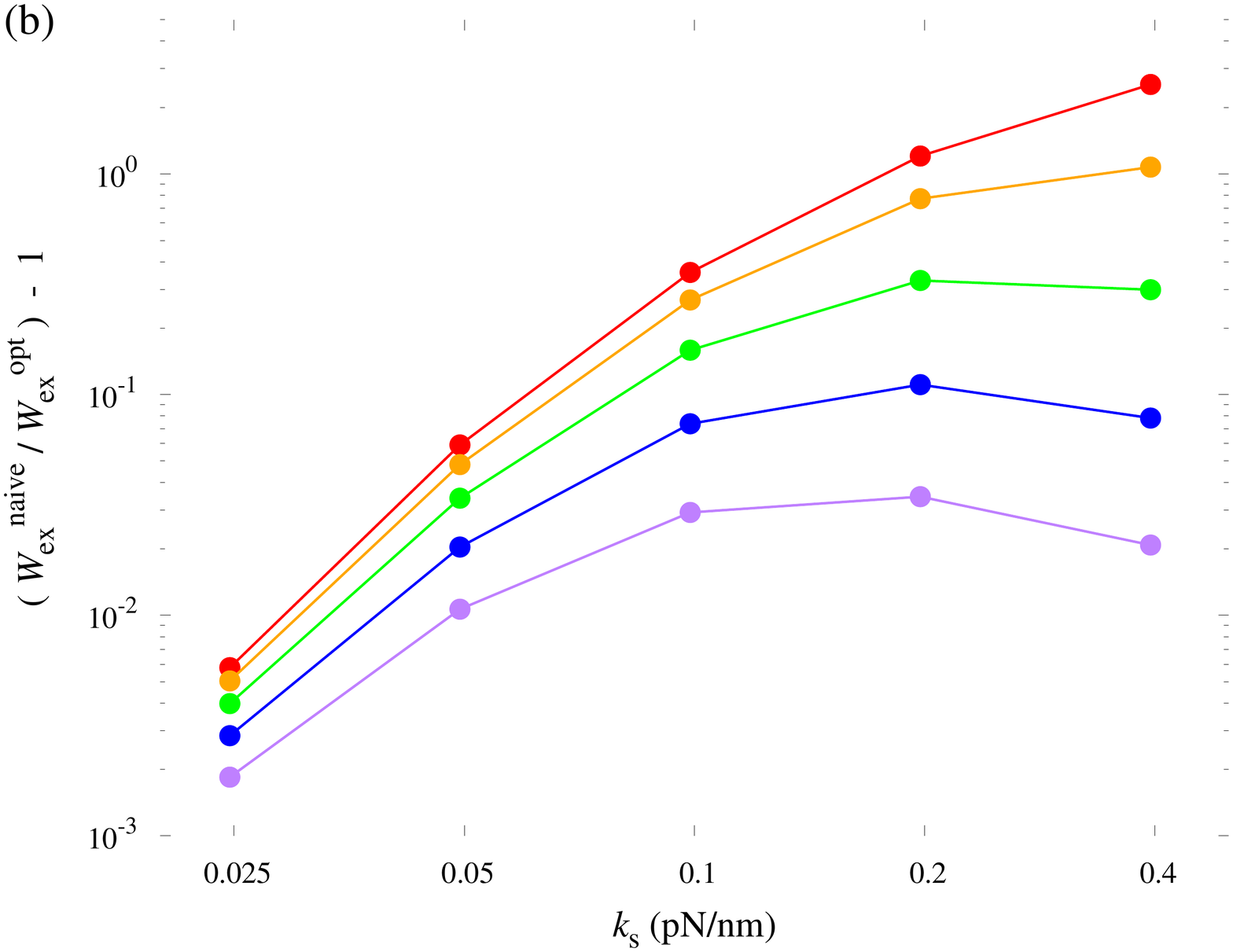}
\caption{(a) Naive excess work, calculated from numerical simulations. (b) Ratio of excess works [Eq.~\eqref{eq:workRatio}] for the naive (constant-velocity) and optimal (minimum-dissipation) protocols, estimated from the linear response approximation. $\kS$ varies within a curve and $\bDEdM$ varies across curves.}
\label{fig:ratioDiss}
\end{figure}

Integrating the excess power gives the excess work for the entire protocol. A simple derivation (Appendix \ref{sec:workRatio}) shows that in the linear response regime the ratio of average excess works in the naive and optimal protocols takes a simple form, the ratio of the average friction coefficient to the square of the mean square-root friction coefficient:
\be
\label{eq:workRatio}
\f{\Wex^{\textrm{naive}}}{\Wex^{\textrm{opt}}} = \f{\overline{\zeta}}{\overline{\zeta^{1/2}}^2} \ .
\ee
The overbar represents an average over all control parameter points in the protocol, $\overline{g} \equiv \int \md\lambda \, g(\lambda) / \int \md \lambda \,$. 
This excess work ratio is independent of the protocol time. 
Jensen's inequality~\cite{Cover:2006:Book} and the concavity of the square root imply that the ratio in Eq.~\eqref{eq:workRatio} is no less than unity.  
In the examined parameter range, numerics show that this excess work coefficient reaches as high as 2.5 (Fig.~\ref{fig:ratioDiss}). 

\s{Conclusion}
Using this approximate linear response framework to predict nonequilibrium properties (the excess power) from equilibrium properties (the generalized friction coefficient, composed of equilibrium force variance and relaxation time), we arrive at a picture of the qualitative nature of this generalized friction and hence optimal driving in a potential that is a model for many activated biomolecular processes. 
The intuitive takeaway is that to minimize energy expended to drive a system over a significant energetic barrier in a limited amount of time and hence out of equilibrium, one should rapidly bring the system near the barrier, then reserve most of the available time to sit near the barrier, giving thermal fluctuations the maximum available time to stochastically boost the system over the barrier `for free.'  

We have established in a simple model system that this approximation is accurate for constant-velocity protocols in a parameter regime representing single-molecule force-extension experiments on nucleic acid hairpins. 
Moreover, minimal excess work protocols, which are not in general constant velocity in the control parameter~\cite{Sivak:2012:PhysRevLett:b}, remain closer to equilibrium than naive protocols, and thus are more likely to match the theoretical approximations. 
The generalized friction coefficient differs by orders of magnitude across even modest free energy barriers, and hence the optimal protocols save significant energy expenditure compared to naive ones.  

Minimizing the excess work during nonequilibrium experiments and simulations would yield significant benefits, as protocols producing less excess work require fewer repetitions to achieve a given statistical precision~\cite{Gore:2003:PNAS}. This study suggests a method to do just this: initial equilibrium sampling at equally-spaced points in control parameter space to estimate the equilibrium fluctuations and relaxation time for corresponding control parameter values, followed by inference of an optimal control parameter protocol~\cite{Crooks:2007:PhysRevLett,Shenfeld:2009:PhysRevE,Grosse:2013:NIPS}. 

$\F1$ ATP synthase can be experimentally  driven in a similar fashion~\cite{HironoHara:2005:PNAS}, where the time-varying quadratic potential is a magnetic tweezers, and rotation of the tweezers drives $\F1$ over a succession of energetic barriers separating its various metastable states~\cite{Okuno:2011:Biochem}.  With sufficient separation between the barriers, the minimum-dissipation rotational protocol is a sequence of single-barrier optimal protocols, suggesting a principle (that depends on the heights of energy barriers) for efficient energy transmission from $\Fo$ to $\F1$ subunits of ATP synthase.

\begin{acknowledgments}
The authors thank Leonid Chindelevitch (SFU Computing Science), Bingyun Sun (SFU Chemistry), and Aliakbar Mehdizadeh, Steven J.\ Large, and Alzbeta Medvedova (SFU Physics) for insightful comments on the paper. 
This work was supported by a Natural Sciences and Engineering Research Council of Canada (NSERC) Discovery Grant (D.A.S.) and by U.S. Army Research Laboratory and the U.S. Army Research Office under Contract No.\ W911NF-13-1-0390~(G.E.C.). This research was enabled in part by support provided by WestGrid (www.westgrid.ca) and Compute Canada Calcul Canada (www.computecanada.ca).
\end{acknowledgments}

\appendix

\s{Thermodynamic divergence and thermodynamic length \label{sec:TDTL}}
The Fisher information~\cite{Rao:1945:BullCalcuttaMath,Cover:2006:Book} \be \FI(\xS) \equiv \l\la \l( \f{\pa \ln \pi(x\mid \xS)}{\pa \xS} \r)^2 \r\ra_{\xS} \ee  is proportional to the force variance $\la \delta f^2 \ra_{\xS}$~\cite{ Crooks:2007:PhysRevLett,Sivak:2012:PhysRevLett:b}. 
The thermodynamic divergence, 
\begin{equation}
{\mathcal J}_D^{\lambda_{\textrm{i}} \rightarrow \lambda_{\textrm{f}}} \equiv (\tF - \tI) \int_{\tI}^{\tF} \l( \f{\md\lambda(t)}{\md t} \r)^2 \la \delta f^2 \ra_{\lambda(t)} \ \md t \ ,
\end{equation}
has a relatively simple expression for our system: 
\begin{widetext}
\begin{align}
{\mathcal J}_D^{\xS^{\textrm i} \rightarrow \xS^{\textrm f}} &= \beta \kC \l\{\f{\kS}{\kW}(\xS^{\textrm f}-\xS^{\textrm i}) + \f{2\sinh[\beta \kC\xMin(\xS^{\textrm f}-\xS^{\textrm i})]}{\cosh[\beta \kC\xMin(\xS^{\textrm f}-\xS^{\textrm i})] + \cosh[\beta(\kC\xMin\{\xS^{\textrm f}+\xS^{\textrm i}\} - \eOffset)]}\r\} \ .
\end{align}
\end{widetext}
The first term is proportional to the length of the integration path and reflects the constant term in the expression for force variance [Eq.~\eqref{eq:forceVar}].  For a given integration distance $\Delta x \equiv \xS^{\textrm f}-\xS^{\textrm i}$, the divergence is maximized when the start and end points are equally distant from the variance maximum, $\xS = \eOffset/(2\kC\xMin)$. Like for the force variance, the energy offset $\eOffset$ simply shifts the location of maximal thermodynamic divergence. 

The thermodynamic length 
\begin{equation}
{\mathcal L}^{\lambda_{\textrm i} \rightarrow \lambda_{\textrm f}} \equiv \int_{t_{\textrm i}}^{t_{\textrm f}} \f{\md\lambda(t)}{\md t} \sqrt{ \la \delta f^2 \ra_{\lambda(t)}} \ \md t 
\end{equation}
is a lower bound on dissipation along any protocol in a given time interval between two thermodynamic states~\cite{Crooks:2007:PhysRevLett} and also admits an analytic expression for this system: 
\begin{subequations}
\begin{align}
{\mathcal L}^{\xS^{\textrm i} \rightarrow \xS^{\textrm f}} &= l(\xS^{\textrm f}) - l(\xS^{\textrm i}) \\
l(\xS) &\equiv \tan^{-1} \f{\sinh\beta\l(\kC\xMin \xS-\half \eOffset\r)}{\sqrt{1 + \f{1+\f{\kS}{\kW}}{\beta\kW \xMin^2} \cosh^2\beta\l(\kC\xMin \xS - \half \eOffset\r)}} \\
&\hspace{-2em} + \sqrt{\f{1+\f{\kS}{\kW}}{\beta\kW \xMin^2}} \  \sinh^{-1}\f{\sinh\beta\l(\kC\xMin \xS - \half \eOffset\r)}{\sqrt{1 + \f{\beta\kW \xMin^2}{1+\f{\kS}{\kW}}}} \notag \ .
\end{align}
\end{subequations}

\s{Ratio of naive and optimal excess works \label{sec:workRatio}}
In this linear response framework, the excess work is the time integral of the friction coefficient times the square of the control parameter velocity.  When only a discrete set of $N$ equally-spaced friction coefficients are known and a piecewise constant-velocity protocol is applied over the total range $\Delta \lambda \equiv \lambda_{\textrm f}-\lambda_{\textrm i}$ (for initial and final control parameter values $\lambda_{\textrm{i}}$ and $\lambda_{\textrm{f}}$, respectively), this integral is approximated by the discrete sum over the constant-velocity segments:  
\begin{subequations}\begin{align}
\Wex &= \int \md t \l( \f{\md \lambda}{\md t} \r)^2 \zeta\bigl[\lambda(t)\bigr] \\
&\approx \sum_j \Delta t_j \l( \f{\md \lambda}{\md t}\Big|_{t_j} \r)^2 \zeta\bigl[\lambda(t_j)\bigr] \\
&= \sum_j \f{\f{\Delta \lambda}{N}}{\f{\md \lambda}{\md t}\Big|_{t_j}} \l( \f{\md \lambda}{\md t}\Big|_{t_j} \r)^2 \zeta_j \\
&= \f{\Delta \lambda}{N} \sum_j \f{\md \lambda}{\md t}\Big|_{t_j} \zeta_j 
 \ .
\end{align}\end{subequations}
For the naive protocol, the control parameter velocity is constant and hence the excess work is proportional to the average friction coefficient:  
\be
\Wex^{\textrm{naive}} = \f{(\Delta \lambda)^2}{N\Delta t} \sum_j \zeta_j 
= \f{(\Delta \lambda)^2}{\Delta t} \overline{\zeta} \ .
\ee
For the optimal protocol, whose control parameter velocity $\md\lambda^{\text{opt}} / \md t$ is proportional to the inverse square root of the friction coefficient, the proportionality constant $A$ is found by requiring that the protocol traverse $\Delta \lambda$ in allotted time $\Delta t$:
\begin{subequations}\begin{align}
\Delta t &= \sum_j \f{\f{\Delta \lambda}{N}}{\f{\md \lambda^{\mathrlap{\text{opt}}}}{\md t}\quad\Big|_{t_j}} \\
&= \sum_j \f{\f{\Delta \lambda}{N}}{A \zeta_i^{-1/2}} \\
&= \f{\Delta \lambda}{A} \f{\sum_i \zeta_i^{1/2}}{N} \\
&= \f{\Delta \lambda}{A} \overline{\zeta^{1/2}} \\
A &= \f{\Delta \lambda}{\Delta t} \overline{\zeta^{1/2}} \ . 
\end{align}\end{subequations}
Thus the optimal protocol requires average excess work
\begin{subequations}\begin{align}
\Wex^{\textrm{opt}} &= \f{\Delta \lambda}{N} \sum_i (A \zeta_i^{-1/2}) \zeta_i  \\
&= \f{(\Delta \lambda)^2}{N\Delta t} \overline{\zeta^{1/2}} \sum_i \zeta_i^{1/2}  \\
&= \f{(\Delta \lambda)^2}{\Delta t} \overline{\zeta^{1/2}}^2 \ , 
\end{align}\end{subequations}
proportional to the square of the mean square root friction coefficient. The ratio of naive and optimal excess works cancels the identical prefactors $(\Delta \lambda)^2/\Delta t$, leaving
\be
\f{\Wex^{\textrm{naive}}}{\Wex^{\textrm{opt}}} = \f{\overline{\zeta}}{\overline{\zeta^{1/2}}^2} \ .
\ee
~

%\bibliography{doubleWell}

\end{document}